\documentclass[12pt]{iopart}
\usepackage{graphicx}
\usepackage{color}

\begin{document}

\title[]{Highly persistent spin textures with giant tunable spin splitting in the two-dimensional germanium monochalcogenides}

\author{Moh. Adhib Ulil Absor$^{1,*}$, Yusuf Faishol$^{1}$, Muhammad Anshory$^{1}$, Iman Santoso$^{1}$, Sholihun$^{1}$, Harsojo$^{1}$, and Fumiyuki Ishii$^{2}$}
\address{$^{1}$Department of Physics, Universitas Gadjah Mada, Sekip Utara BLS 21 Yogyakarta 55281, Indonesia.}
\address{$^{2}$Nanomaterials Research Institute, Kanazawa University, 920-1192, Kanazawa, Japan.}
\ead{$^{*}$adib@ugm.ac.id}

\vspace{10pt}
\begin{indented}
\item[]February 2021
\end{indented}

\begin{abstract}
The ability to control the spin textures in semiconductors is a fundamental step toward novel spintronic devices, while seeking desirable materials exhibiting persistent spin texture (PST) remains a key challenge. The PST is the property of materials preserving a unidirectional spin orientation in the momentum space, which has been predicted to support an extraordinarily long spin lifetime of carriers. Herein, by using first-principles density functional theory calculations, we report the emergence of the PST in the two-dimensional (2D) germanium monochalcogenides (GeMC). By considering two stable formation of the 2D GeMC, namely the pure Ge$X$ and Janus Ge$_{2}XY$ monolayers ($X,Y$ = S, Se, and Te), we observed the PST around the valence band maximum where the spin orientation is enforced by the lower point group symmetry of the crystal. In the case of the pure Ge$X$ monolayers, we found that the PST is characterized by fully out-of-plane spin orientation protected by $C_{2v}$ point group, while the canted PST in the $y-z$ plane is observed in the case of the Janus Ge$_{2}XY$ monolayers due to the lowering symmetry into $C_s$ point group. More importantly, we find large spin-orbit coupling (SOC) parameters in which the PST sustains, which could be effectively tuned by in-plane strain. The large SOC parameter observed in the present systems leads to the small wavelength of the spatially periodic mode of the spin polarization, which is promising for realization of the short spin channel in the spin Hall transistor devices.
\end{abstract}

%
%
%
%
%

\section{Introduction}

Finding novel materials with strong spin-orbit coupling (SOC) has been one of the important research themes in the field of spintronics \cite{Manchon}. The large SOC effect is indispensable in spintronics since it could effectively manipulate the spin of electron electrically, which plays a central role in many intriguing phenomena such as spin relaxation \cite{Fabian, Averkiev}, hidden spin polarization effect \cite{Zhang2014, Yuan2019, Liu2019, Absor2021}, spin Hall effect \cite{Qi}, spin galvanic effect \cite{Ganichev}, and spin ballistic transport \cite{Lu}. In a system lacking an inversion center, the SOC induces momentum-dependent spin-orbit field (SOF) lifting Kramer's spin degeneracy and leading to a complex $k$-dependent spin texture of the electronic bands through the so-called Rashba\cite{Rashba} and Dresselhaus \cite{Dress} effects. In particular, the Rashba spin texture can be manipulated electrically to produce non-equilibrium spin polarization \cite{Nitta, Kuhlen}, which has potential application as spin-field effect transistor (SFET)  \cite{Datta, Chuang}. Although the large SOC is beneficial for spintronic devices, it is also known to induce the undesired effect of causing spin decoherence. In a diffusive transport regime, the momentum-dependent SOF induces spin randomization by a process known as the Dyakonov-Perel spin relaxation \cite{Dyakonov}, which significantly reduces the spin lifetime, and hence limits the performance of the spintronics functionality. 

The problem of the spin dephasing by the SOC can be eliminated by designing the materials to exhibit unidirectional SOF. Here, the spin texture is enforced to be uniform and independent of the electron momentum, called the persistent spin texture (PST) \cite{Schliemann, Bernevig, SchliemannJ}, arising when the linear Rashba and Dresselhaus contributions compensate each other. Such peculiar spin textures lead to a spatially periodic mode of the spin polarization in the crystal known as persistent spin helix (PSH) mode, enabling long-range spin transport without dissipation \cite{Bernevig, Altmann}. The PST has been previously observed on [001]-oriented semiconductors quantum well (QW) having an equal the Rashba and Dresselhaus SOC parameters \cite{Koralek2009, Walser2012, Kohda, Sasaki2014}, or on [110]-oriented semiconductor QW \cite{Chen} described by the [110] Dreseelhauss model \cite{Bernevig}. Similar to the [110]-oriented QW, the PST has also been reported on strained LaAlO$_{3}$/SrTiO$_{3}$ interface \cite{Yamaguchi_2017}. Although achieving the PST requires controlling the Rashba and Dresselhaus SOC parameters, it is technically non-trivial since both the parameters are material dependent. This has triggered much attention to finding novel systems where the PST can be observed intrinsically. 

Recently, the concept of the PST has been developed in a more general way by enforcing the symmetry of the crystal rather than fine-tuning the SOC parameters.  For instant, the PST protected by nonsymmorphic space group has been proposed, as recently reported on various three-dimensional (3D) bulk systems such as BiInO$_{3}$ with $Pna2_{1}$ space group \cite{Tao2018}, CsBiNb$_{2}$O$_{7}$ with $P2_{1}am$ space group \cite{Autieri2019}, and Ag$_{2}$Se with $P222_{1}$ space group \cite{Shi2020}. Moreover, the symmetry-protected PST with purely cubic spin splitting has been predicted in the 3D bulk materials crystallizing in the $\bar{6}m2$ and $\bar{6}$ point groups, as found on the 3D bulk Ge$_{3}$Pb$_{5}$O$_{11}$, Pb$_{7}$Br$_{2}$F$_{12}$, and Pb$_{7}$Cl$_{2}$F$_{12}$ compounds \cite{Zhao2020}. Furthermore, the canted PST has been reported in wurtzite ZnO [10$\bar{1}$0] surface having $C_{s}$ point group symmetry imposed by the non-polar direction of the surface \cite{Absor2015}. In addition, the symmetry-enforced PST has also been observed in several two-dimensional (2D) systems, although it is still very rarely discovered. For the best of our knowledge, only few classes of the 2D materials that has been reported to support the PST including WO$_{2}$Cl$_{2}$ monolayer with $Pmc2_{1}$ space group \cite{Ai2019} and various group-IV monochalcogenide monolayers with $Pmn2_{1}$ space group such as SnSe \cite{Absor2019a,Anshory2020} and SnTe monolayers \cite{Absor2019b,Lee2020}. More recently, the PST induced by the lower symmetry of the structure has been predicted on several 2D transition metal dichalcogenides with the line defect \cite{Li2019, Absor2020}. 

Although the PST has been widely studied in the 3D bulk and QW systems, the search for the ultra-thin 2D materials supporting the PST has lately been very demanding of attention because of their potential for miniaturization spintronic devices \cite{Absor2019b, Anshory2020, Lee2020}. In this work, we predict the emergence of the PST in the 2D germanium monochalcogenides (GeMC) monolayers by using first-principles density functional theory calculations. We have considered two stable formations of the 2D GeMC monolayers, namely the pure Ge$X$ and Janus Ge$_{2}XY$ monolayers ($X, Y$ = S, Se, and Te), and found that the PST is observed around the valence band maximum where the spin orientation is enforced by the lower point group symmetry of the crystal. In the case of the pure Ge$X$ monolayers, we found that the PST is characterized by the fully out-of-plane spin orientation, which is protected by the $C_{2v}$ point group. On the other hand, the PST is canted in the $y-z$ plane for the case of the Janus Ge$_{2}XY$ monolayers, which is due to the lowering symmetry into $C_s$ point group. More interestingly, we identified large SOC parameters in the spin-split bands where the PST maintains, which could be effectively regulated by applying in-plane strain. The observed large SOC parameter in the present systems results in that the small wavelength of the spatially periodic mode of the spin polarization is achieved. Thus, we proposed the present system as a short spin channel in the spin Hall transistor devices.  

\begin{figure}[h!]
	\centering	\includegraphics[width=0.7\linewidth]{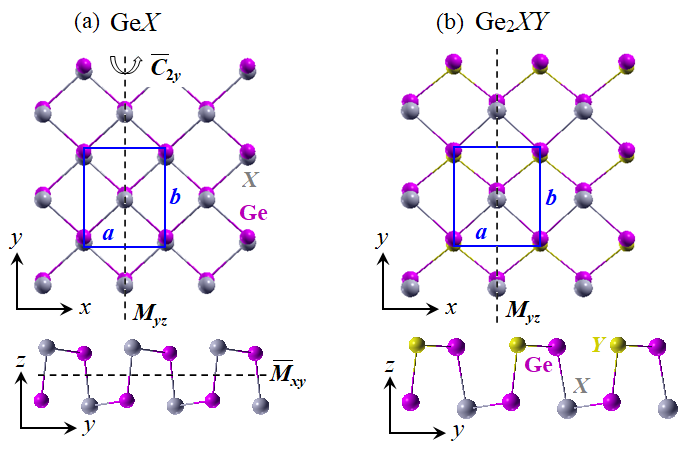}
	\caption{Crystal structure corresponding to the symmetry operations for: (a) pure 2D Ge$X$ and (b) Janus Ge$_{2}XY$ monolayers. The unit cell of the crystal is indicated by blue lines characterized by $a$ and $b$ lattice parameters in the $x$ and $y$ directions, respectively.}
	\label{fig:1}
\end{figure}

\section{Model and Computational Details}

The 2D GeMC monolayers crystallize in a black phosphorene-type structures \cite{Appelbaum2016}, forming two different stable structures, called as the pure Ge$X$ and Janus Ge$_{2}XY$ monolayers [Fig. 1(a)-(b)]. The pure Ge$X$ monolayers have $C^{7}_{2v}$ or $Pmn2_{1}$ space group characterized by four symmetry operations in the crystal lattice [Fig. 1(a)]: (i) identity operation $E$; (ii) twofold screw rotation $\bar{C}_{2y}$ (twofold rotation around the $y$ axis, $C_{2y}$, followed by translation of $\tau=a/2,b/2$), where $a$ and $b$ is the lattice parameters along $\vec{a}$ and $\vec{b}$ directions, respectively; (iii) glide reflection $\bar{M}_{xy}$ (reflection with respect to the $xy$ plane followed by translation $\tau$); and (iv) reflection $M_{yz}$ with respect to the $yz$ plane. By replacing one of the $X$ atoms in the Ge$X$ monolayers with the $Y$ atoms, we get the Janus Ge$_{2}XY$ monolayers [Fig. 1(b)]. Here, only the $M_{yz}$ mirror symmetry survives, thus lowering the symmetry into $C_{s}$ point group.

We have performed non-collinear first-principles DFT calculations using the OpenMX code \cite{Openmx, Ozaki, Ozakikino, Ozakikinoa}, based on the norm-conserving pseudopotentials and optimized pseudoatomic localized basis functions \cite{Troullier}. The energy cutoff of 350 Ry was used for charge density. The exchange-correlation functional was treated within generalized gradient approximation by Perdew, Burke, and Ernzerhof (GGA-PBE)\cite{gga_pbe, Kohn}. We used the $12\times10\times1$ $k$-point mesh to discretize the first Brillouin zone. The Pseudo-atomic basis functions of $s^{2}p^{2}d^{2}f^{1}$ along with the norm-conserving pseudopotentials were used from the OpenMX library. We used the $j$-dependent pseudo potentials where the SOC interaction was included self consistently in all calculations \cite{Theurich}. 

We calculated the spin textures by deducing the spin vector components ($S_{x}$, $S_{y}$, $S_{z}$) in the reciprocal lattice vector $\vec{k}$ from the spin density matrix \cite{Kotaka_2013}. The spin density matrix, $P_{\sigma \sigma^{'}}(\vec{k},\mu)$, were calculated using the following relation,  
\begin{equation}
\label{1}
P_{\sigma \sigma^{'}}(\vec{k},\mu)=\int \Psi^{\sigma}_{\mu}(\vec{r},\vec{k})\Psi^{\sigma^{'}}_{\mu}(\vec{r},\vec{k}) d\vec{r},
\end{equation}
where $\Psi^{\sigma}_{\mu}(\vec{r},\vec{k})$ is the spinor Bloch wave function obtained from the non-collinear first-principles calculations.

\begin{table}[ht!]
\caption{The optimized lattice parameters [$a$ (\AA), $b$ (\AA)], structure anisotropy factor, $\kappa$, defined as $\kappa=(a-b)/(a+b)$, the formation energy, $E_{f}$ (eV), and the band gap, $E_{g}$ (eV), obtained for the 2D GeMC monolayers. The star (*) indicates the semiconductor with direct bandgap.} 
\begin{indented}
\item[]
\begin{tabular}{c c c c c c } 
\hline\hline 
GeC Monolayers & $a$ (\AA)  & $b$ (\AA) &  $\kappa$ &  $E_{f}$ (eV) &  $E_{g}$ (eV)\\ 
\hline 
 GeS &  3.68   &    4.40  &  0.09 &     &  1.45    \\     
 GeSe&  3.99   &    4.39  &  0.05  &     &  1.10$^{*}$    \\		 
 GeTe &  4.27   &    4.47 &  0.02  &   &  0.92   \\     
 Ge$_{2}$SSe&  3.84   &   4.47  & 0.08  & 0.03  & 1.32  \\            
 Ge$_{2}$STe&  4.03   &    4.53  &  0.06  & 0.15    & 0.97  \\            
 Ge$_{2}$SeTe&  4.14   &    4.47  & 0.04  & 0.05  & 0.88   \\            
\hline\hline 
\end{tabular}
\end{indented}
\label{table:Table 1} 
\end{table}

To avoid artificial interactions between periodic images created by the periodic boundary condition, we used periodic slab model for both the Ge$X$ and Janus Ge$_{2}XY$ monolayers with a sufficiently large vacuum layer (20 \AA) in the non-periodic direction. The lattice and positions of the atoms were optimized until the Hellmann-Feynman force components acting on each atom was less than $10^{-3}$ eV\AA\, where the energy convergence criterion was set to $10^{-9}$ eV. In the case of the Janus Ge$_{2}XY$ monolayers, the energetic stability of the structure is confirmed by calculating the formation energy, $E_{f}$, through the following relation,
\begin{equation}
\label{2}
E_{f}=E_{tot}[\texttt{Ge}_{2}XY]-\frac{1}{2}E_{tot}[\texttt{Ge}X]-\frac{1}{2}E_{tot}[\texttt{Ge}Y],  
\end{equation}
where $E_{tot}[\texttt{Ge}_{2}XY]$, $E_{tot}[\texttt{Ge}X]$, and $E_{tot}[\texttt{Ge}Y]$ are the total energy of Ge$_{2}XY$, Ge$X$, and Ge$Y$, respectively. The optimized structural-related parameters of the Ge$X$ and Janus Ge$_{2}XY$ monolayers, and the formation energy of the Janus Ge$_{2}XY$ monolayers are summarized in Table 1, and are in a good agreement overall with previously reported data \cite{Absor2019b,Xu_2017,Seixas2020}. 

\begin{figure*}[h!]
	\centering	\includegraphics[width=1.1\linewidth]{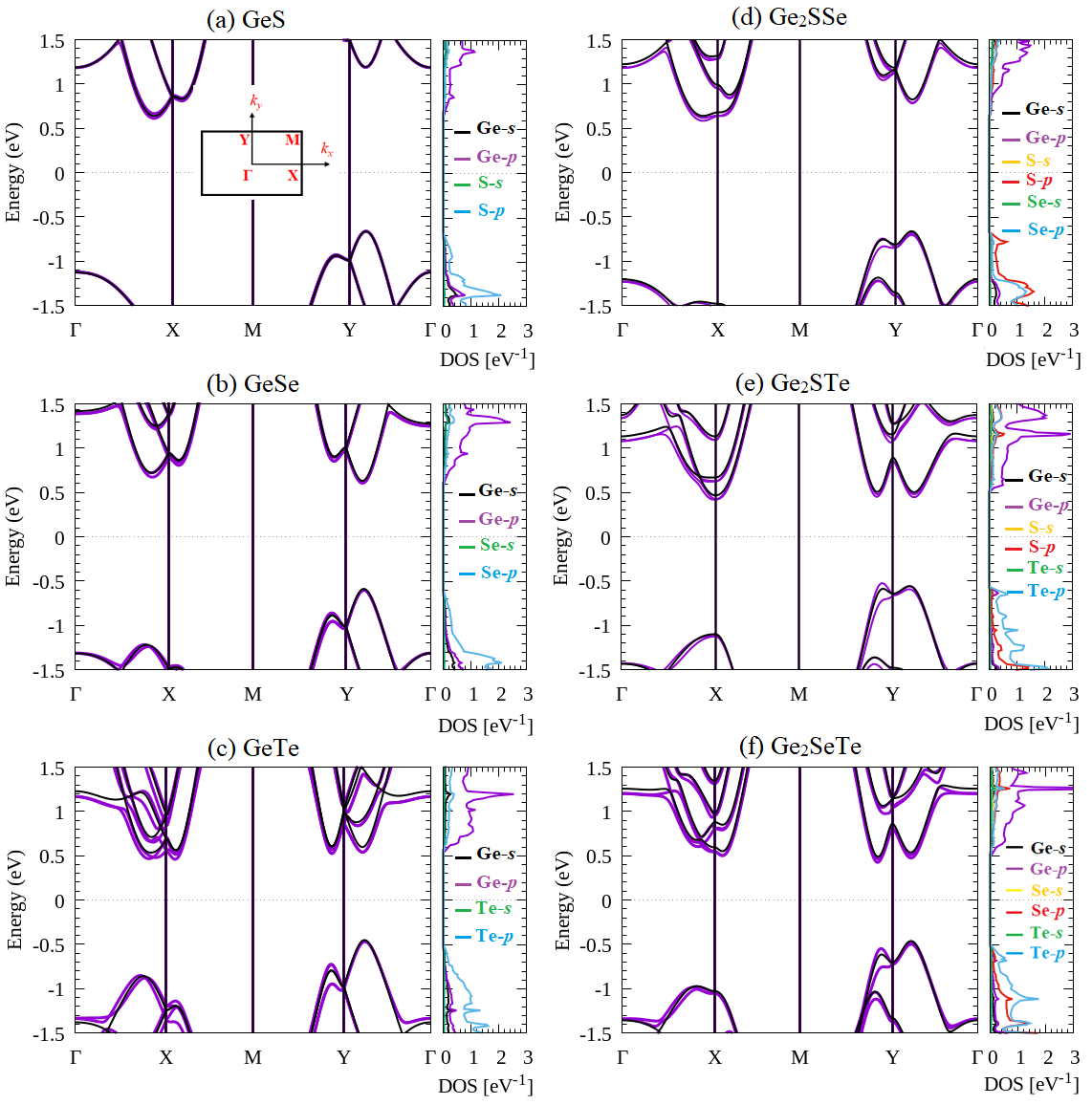}
	\caption{Electronic band structure corresponding to the density of states (DOS) projected to the atomic orbitals for: (a) GeS, (b) GeSe, (c) GeTe, (d) Ge$_{2}$SSe, (e) Ge$_{2}$STe, and (f) Ge$_{2}$SeTe monolayers. For the band structure, the black and pink colours indicate the calculated bands without and with the SOC, respectively. The insert of Fig. 2(a) shows the first Brillouin zone used in the band structure calculations.}
	\label{fig:2}
\end{figure*}

\section{Results and discussion}

Fig. 2 shows the band structure of various 2D GeMC monolayers calculated along the first Brillouin zone (FBZ) [see the insert of Fig. 2(a)] corresponding to their density of states (DOS) projected to the atomic orbitals. Without the SOC, the GeCM monolayers are semiconductors with an indirect band gap, except for the GeSe monolayer having a direct band gap. The calculated band gap is smaller for the heavier materials (GeTe, Ge$_{2}$SeTe) and larger for the lighter materials (GeS, Ge$_{2}$SSe) [see Table 1], which is consistent with previous reports \cite{Absor2019b,Kamal2016,Seixas2020}. The valence band maximum (VBM) is located at the $\Gamma-Y$ line, while the conduction band minimum (CBM) show a different location for the different chalcogen atoms. In the case of the pure Ge$X$ monolayers, the CBM is located at the $\Gamma-X$ line for both the GeS and GeTe monolayers [Figs. 2(a) and 2(c)], while it is located at the $\Gamma-Y$ line for the GeSe monolayer [Fig. 2(b)]. In contrast, for the case of the Janus Ge$_{2}XY$ monolayers, the CBM is located at the $\Gamma-X$ line, $\Gamma$ point, and $M-Y$ line for the Ge$_{2}$SSe, Ge$_{2}$STe, and Ge$_{2}$SeTe monolayers, respectively [Figs. 2(d)-(f)]. Our calculated DOS projected to the atomic orbitals confirmed that the CBM is mainly originated from the contribution of the Ge-$p$ and $X$($Y$)-$s$ orbitals, while the VBM is dominated by the Ge-$s$ and $X$($Y$)-$p$ orbitals. 

Turning the SOC leads to a sizable splitting of the bands due to the broken of the inversion symmetry, which is especially pronounced in the bands around the $Y$ point at the VBM [Fig 2(a)-(f)]. Here, the larger band splitting is identified for the monolayers with heavier elements such as Se and Te atoms. However, we observe the band degeneracy for the $\vec{k}$ along the $\Gamma-Y$ line in the case of the pure Ge$X$ monolayers, which is protected by the $M_{yz}$ and $\bar{M}_{xy}$ symmetry operations [see Appendix A for detail symmetry analysis]. Conversely, this degeneracy is lifted for the case of the Ge$_{2}XY$ monolayers, which is due to the broken of the $\bar{C}_{2y}$ and $\bar{M}_{xy}$ symmetry operations. Since both the the GeTe and Ge$_{2}$SeTe monolayers have the largest band splitting among the members of 2D GeMC monolayers, in the following discussion, we focused on these monolayers as a representative example of the pure Ge$X$ and Ge$_{2}XY$ monolayers, respectively.

\begin{figure*}[h!]
	\centering	\includegraphics[width=1.1\linewidth]{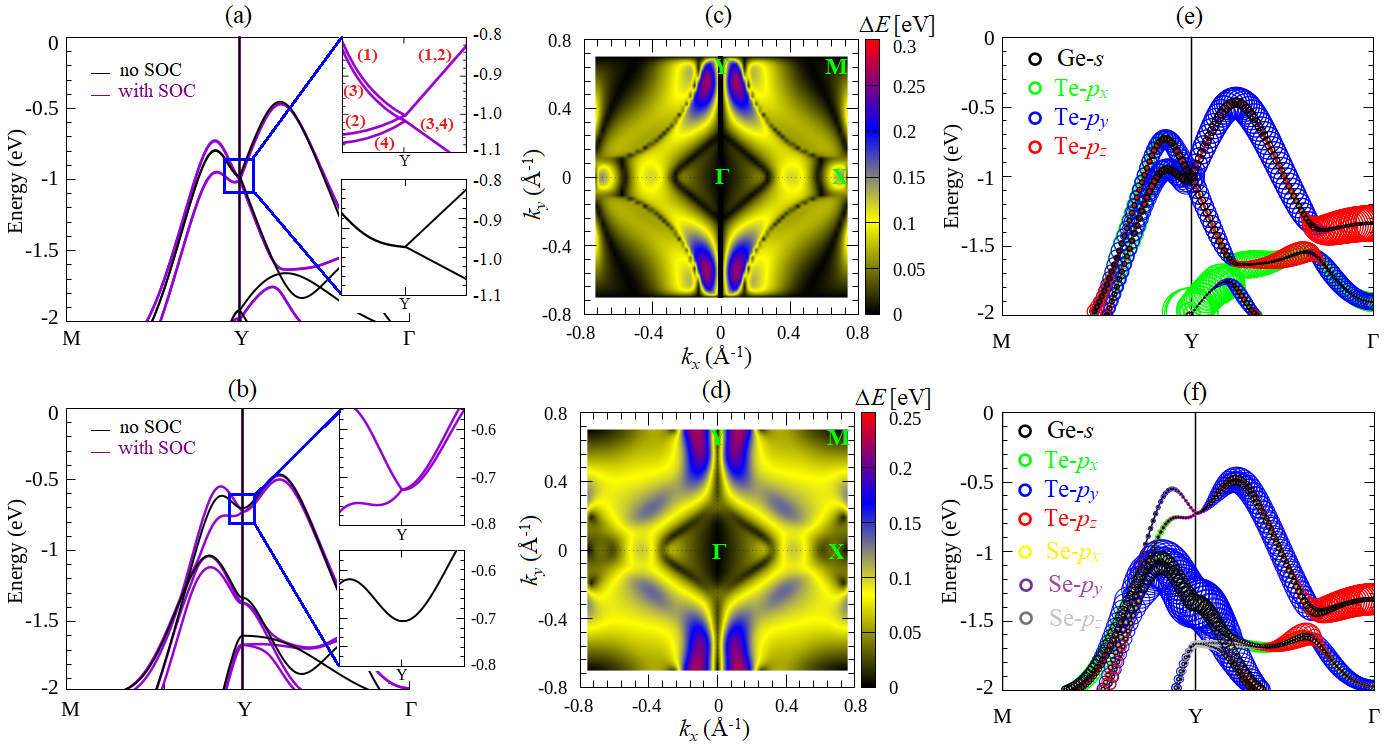}
	\caption{(a)-(b) The valence bands of the GeTe and Ge$_{2}$SeTe monolayers, respectively, calculated around the $Y$ point are presented. The black and red colors show the calculated bands without and with the SOC, respectively. The insert shows the spin-split bands of the valence band maximum (VBM) calculated very close to the $Y$ point where the band indexes are represented by the number (1) to (4) for the first four valence bands. (c)-(d) Spin-splitting energy of the bands around the VBM mapped on the $k$-space in the first Brillouin zone for the GeTe and Ge$_{2}$SeTe monolayers, respectively, are shown. The magnitude of the spin-splitting energy, $\Delta E$, defined as $\Delta E=|E(k,\uparrow)_{(1)}-E(k,\downarrow)_{(2)}|$, where $|E(k,\uparrow)_{(1)}$ and $E(k,\downarrow)_{(2)}$ are the energy bands of the state (1) with up spin and the state (2) with down spin, respectively, is represented by the color scales. (c)-(d) Orbital-resolved electronic band structures calculated around the VBM for the GeTe and Ge$_{2}$SeTe monolayers, respectively, are presented. The radii of the circles reflect the magnitudes of the spectral weight of the particular orbitals to the band.}
	\label{fig:3}
\end{figure*}

Figs. 3(a)-(b) show the valence band of the GeTe and Ge$_{2}$SeTe monolayers, respectively, calculated around the $Y$ point, where the spin-split bands at the VBM are highlighted. Without the SOC, the VBM of the GeTe monolayer exhibits a fourfold degenerate state along the $Y-M$ line but splits into two pair doublets along the $Y-\Gamma$ line [Fig. 3(a)]. When the SOC is introduced, the doublets along the $Y-\Gamma$ line retain protected by the $M_{yz}$ and $\bar{M}_{xy}$ symmetry operations [see Appendix A for detail symmetry analysis]. However, along the $Y-M$ line, the SOC splits the degenerate states into singlet [Fig. 3(a)], resulting in a strongly anisotropic spin splitting around the $Y$ point similar to that observed on various 2D group IV monochalcogenide \cite{Absor2019b, Appelbaum2016, Anshory2020}. In contrast, the VBM along the $Y-M$ and $Y-\Gamma$ lines shows a twofold degenerate state for the case of the Ge$_{2}$SeTe monolayer, which splits into singlet when the SOC is taken into account except for the $Y$ point due to time reversibility [Fig. 3(b)].

To quantify the spin-split bands, we show in Figs. 3(c)-(d) the calculated spin-splitting energy of the VBM mapped along the entire FBZ for GeTe and Ge$_{2}$SeTe monolayers, respectively. As expected, along the $Y-\Gamma$ line, zero splitting energy is observed for the case of the GeTe monolayer [Fig. 3(c)], while the substantially small splitting energy (up to $0.02$ eV) is found in the case of the Ge$_{2}$SeTe monolayer [Fig. 3(d)]. On the other hand, the large splitting energy up to 0.28 eV (0.23 eV) is achieved along the $Y-M$ line at the GeTe (Ge$_{2}$SeTe) monolayer. By calculating orbitals-resolved projected to bands at near the VBM around the $Y$ point, we clarified that the large splitting along the $Y-M$ line is mainly contributed from the strong hybridization between the $s$ orbital of the Ge atom and the $p_{y}$ orbital of the chalcogen atoms (Te-$p_{y}$ orbital in the case of the GeTe monolayer and mixing between Se-$p_{y}$ and Te-$p_{y}$ orbitals for the case of the Ge$_{2}$SeTe monolayer) [Fig. 3(e)-(f)]. Remarkably, the calculated splitting energies in both the GeTe and Ge$_{2}$SeTe monolayers are comparable with that observed on several 2D materials including the pure and Janus transition metal dichalcogenide [0.15 eV - 0.55 eV] \cite{Zhu2011, Yao2017, Absor2016, Absor2017}.

To further demonstrate the nature of the spin splitting, we show in Figs. 4(a) and 4(b) the calculated results of the spin textures around the $Y$ point near the VBM for the GeTe and Ge$_{2}$SeTe monolayers, respectively. We also highlight the spin textures by providing the spin-resolved projected to the bands around the VBM along the $M-Y-\Gamma$ lines as depicted in Figs. 4(c) and 4(d). It is clearly seen that the spin textures of the GeTe monolayer are characterized by fully out-of-plane spin components $S_{z}$, while the in-plane spin components ($S_{x}$, $S_{y}$) are almost zero [Figs. 4(a) and 4(c)]. These spin textures are switched from $S_{z}$ to $-S_{z}$ spin components when crossing at the $k_{x}=0$ along $M-Y$ ($k_{x}$) line [Figs. 4(a)]. We noted here that the switching of the $S_{z}$ spin polarization is also identified at $k_{y}=0.52$ \AA$^{-1}$ along the $Y-\Gamma$ line [Fig. 4(c)], which is due to the equal population between the $S_{z}$ and $-S_{z}$ spin-polarized states in the degenerate bands [see Appendix B for the calculated ensemble-average value of the $S_{z}$ spin components along the $Y-\Gamma$ line]. These features of the spin textures are notably different either from the Rashba \cite{Rashba} and Dresselhaus \cite{Dress} spin textures. Moreover, these typical spin textures gives rise to the so-called persistent spin textures (PST) \cite{Schliemann, Bernevig, SchliemannJ}, which is consistent with that described by the [110] Dresselhaus model in a [110]-oriented semiconductor QW \cite{Bernevig,Chen} and similar to that recently reported on various 2D materials such as WO$_{2}$Cl$_{2}$ \cite{Ai2019}, SnSe \cite{Absor2019a,Anshory2020} and SnTe monolayers \cite{Absor2019b,Lee2020}. The emergence of the PST preserves in the case of the Ge$_{2}$SeTe monolayer but it shows the different features, i.e., the spin textures are mainly characterized by $S_{z}$ and $S_{y}$ spin components, except at $k_{x}=0$ (along the $\Gamma-Y$ line) where the $S_{x}$ spin component retains [Figs. 4(b) and 4(d)]. Accordingly, quasi-one-dimensional spin textures are observed, which is uniformly tilted from the out-of-plane $z$- to the in-plane $y$-direction at $k_{x}\neq 0$, forming a canted PST in the $y-z$ plane similar to that observed on ZnO (10$\bar{1}$0) surface \cite{Absor2015}. The existence of the PST in both the GeTe and Ge$_{2}$SeTe monolayers is expected to induce a unidirectional SOF, protecting the spins from decoherence through suppressing the Dyakonov-Perel spin-relaxation mechanism \cite{Dyakonov}. This is highly beneficial to support ultimately long spin-lifetime of carriers \cite{Altmann}, which is promising for the realization of an efficient spintronics device. 

\begin{figure*}
	\centering	\includegraphics[width=1.0\linewidth]{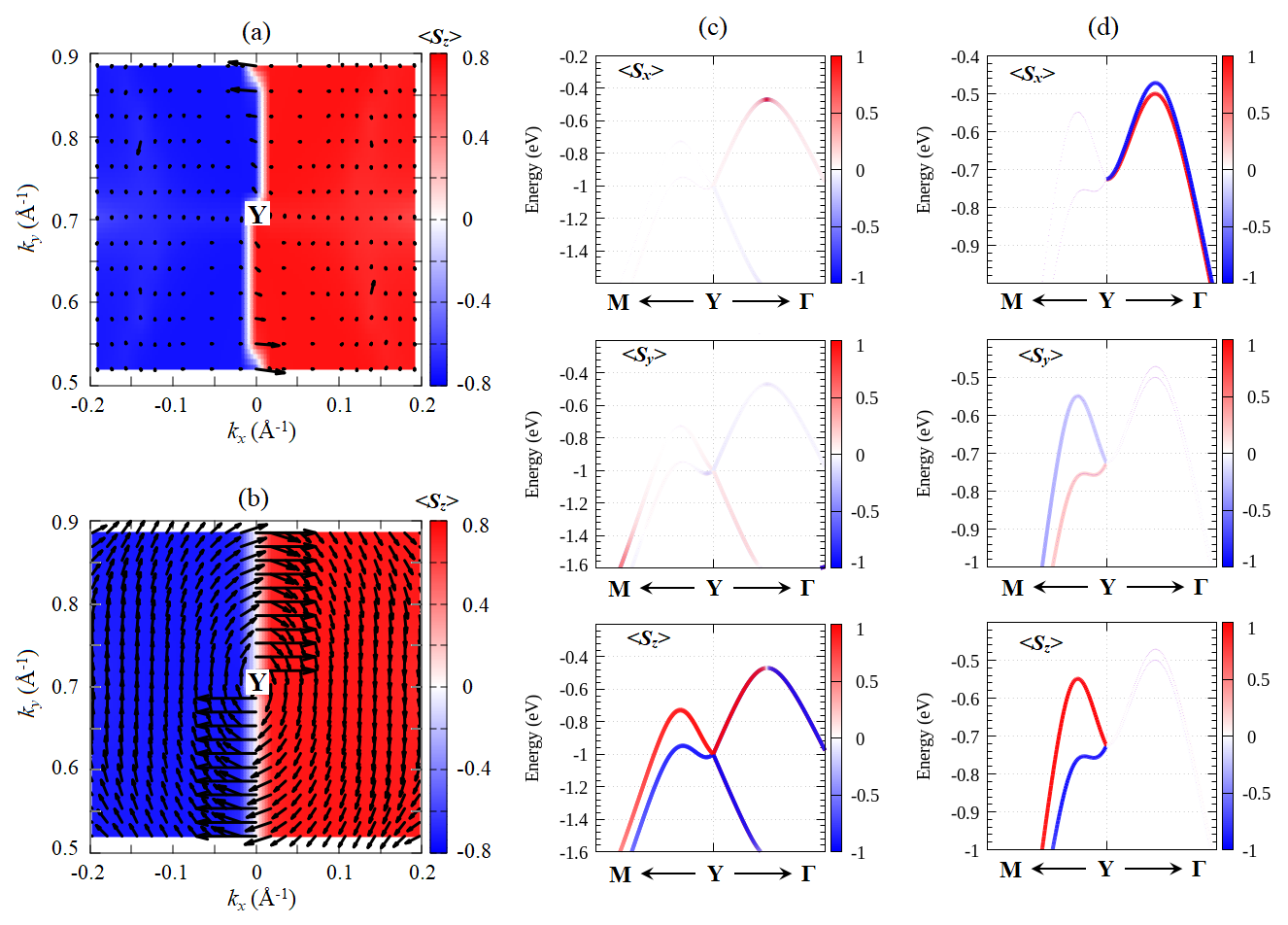}
	\caption{The spin textures around the $Y$ point at the outer branch of the spin-split bands at the VBM calculated for (a) GeTe and (b) Ge$_{2}$SeTe monolayers. In Fig. 4(a)-(b), the arrows show the in-plane components of the spin, while the color indicate the out-of-plane component of the spins, $S_{z}$. Spin-resolved projected to the spin-split bands at the VBM along the $M-Y-\Gamma$ lines  for: (c) GeTe and (d) Ge$_{2}$SeTe monolayers. The color bars in Fig. 4(c)-(d) represents the expected values of the $S_{x}$,  $S_{y}$, and  $S_{z}$ spin components.}
	\label{fig:4}
\end{figure*}

The physical origin of the band splitting and spin textures observed around the $Y$ point can be clarified by using a simple Hamiltonian model in the presence of SOC derived from the symmetry of the wave vector $k$. Since the GeTe monolayer possesses $C_{2v}$ point group symmetry at $Y$ point, the symmetry adapted $\vec{k}\cdot\vec{p}$ Hamiltonian can be written in the linear term of the $k$ as
\begin{equation}
\label{3} 
H_{Y}^{C_{2v}}=E_{0}(k)+\alpha k_{x}\sigma_{z}, 
\end{equation}
where $E_{0}(k)=\hbar^{2}[(k_x^2/2m_{x}^{*})+(k_y^2/2m_{y}^{*})]$ is the nearly-free-hole energy, $k_{x}$ and $k_{y}$ are the wave vectors in the $x$- and $y$-directions, respectively, $m_{x}^{*}$ and $m_{y}^{*}$ are the hole effective mass in the $x$- and $y$-directions, respectively, $\sigma_{z}$ is the $z$-component of the Pauli matrices, and $\alpha$ is the SOC parameter in the spin-split bands along the $k_{x}$ direction [details derivation of the Hamiltonian $H_{Y}^{C_{2v}}$, see Appendix C].

\begin{table}[ht!]
\caption{The calculated SOC parameter ($\alpha$) [in eV\AA] and the wavelength of the PSH ($l_{\texttt{PSH}}$) [in nm] for the GeTe and Ge$_{2}$SeTe monolayers at the VBM around the $Y$ point compared with those observed on various PST materials.} 
\begin{indented}
\item[]
\begin{tabular}{c c c c} 
\hline\hline 
  PST Systems & $\alpha$ (eV\AA)  & $l_{PSH}$ (nm)  & Reference \\ 
\hline 
\textbf{GeMC} &   &   &    \\ 
 GeTe &  3.93   &    6.53    &   This work \\
 Ge$_{2}$SeTe&  3.10   &    8.52    &   This work \\
\textbf{Semiconductor QW}    &    &  &   \\ 
GaAs/AlGaAs & (3.5 - 4.9)$\times 10^{-3}$  & (5.0 - 10) $\times10^{3}$& Ref.\cite{Koralek2009,Walser2012} \\               
InAlAs/InGaAs & 1.0 - 2.0 $\times10^{-3}$&      & Ref.\cite{Kohda,Sasaki2014}\\             
Strained LaAlO$_{3}$/SrTiO$_{3}$ & 7.49 $\times10^{-3}$ & 9.8  & Ref.\cite{Yamaguchi_2017}\\
\textbf{Semiconductor Surface}    &    &  &   \\               
ZnO(10-10) surface & 34.78 $\times10^{-3}$& 1.9$\times10^{2}$ & Ref.\cite{Absor2015}\\
\textbf{3D Bulk systems}   &    &  &   \\ 
BiInO$_{3}$ &1.91 & 2.0 & Ref.\cite{Tao2018}\\
CsBiNb$_{2}$O$_{7}$ &0.012 - 0.014 &  & Ref.\cite{Autieri2019}\\
\textbf{2D systems}   &    &  &   \\ 
SnTe & 1.2 - 2.85& 1.82 - 8.8  & Ref.\cite{Absor2019b,Lee2020}  \\      
SnSe & 0.76 - 1.15 &  & Ref.\cite{Anshory2020}\\
Doped SnSe & 1.6 - 1.76 & 1.2 - 1.41 & Ref.\cite{Absor2019a}\\
WO$_{2}$Cl$_{2}$ &  0.9     &      & Ref.\cite{Ai2019}\\
PtSe$_{2}$ with line defect & 0.2 - 1.14 & 6.33 -  28.19 & Ref.\cite{Absor2020}\\
WS$_{2}$ with line defect & 0.14 - 0.26 & 8.56 - 10.18 & Ref.\cite{Li2019}\\
\hline\hline 
\end{tabular}
\end{indented}
\label{table:Table 2} 
\end{table}

We can see clearly that the derived $H_{Y}^{C_{2v}}$ in Eq. (\ref{3}) is only coupled with $k_{x}\sigma_{z}$ term, justifying that the spin textures around the $Y$ point being oriented in the fully out-of-plane $z$-direction. As expected, the out-of-plane PST is achieved, which is consistent-well with the calculated spin textures shown in Figs. 4(a) and 4(c). Moreover, by solving the eigenvalue problem involving the $H_{Y}^{C_{2v}}$, we obtain that the energy dispersion can be expressed as $E^{C_{2v}}_{Y,\pm}(k)=E_{0}(k)\pm\alpha k_{x}$. This fact implies that the bands are lifted along the $Y-M$ line ($k_{x}$) but are degenerated along the $Y-\Gamma$ line ($k_{y}$), which is in agreement with the calculated band dispersion of the VBM around the $Y$ point shown in Fig. 3(a). By fitting  $E^{C_{2v}}_{Y,\pm}(k)$ to the DFT band dispersion of the GeTe monolayer around the $Y$ point at the VBM, we find that the calculated value of $\alpha$ is 3.93 eV\AA, which supports the large spin-splitting energy along the $Y-M$ line ($k_{x}$) in agreement with the DFT results provided in Fig. 3(c). 

A similar analysis can also be applied to explain the band splitting and spin textures in the Ge$_{2}$SeTe monolayer. Here, the lowering symmetry into $C_{s}$ point group leads to the fact that the $\vec{k}\cdot\vec{p}$ Hamiltonian around the $Y$ point can be expressed as 
\begin{equation}
\label{4} 
H_{Y}^{C_{s}}=E_{0}(k)+(\alpha_{1} \sigma_{z}+\alpha_{2} \sigma_{y}) k_{x}+\beta k_{y}\sigma_{x},
\end{equation}
where $\sigma_{x}$, $\sigma_{y}$, $\sigma_{z}$ are  the $x$-, $y$-, and $z$-components of the Pauli matrices, respectively, $\alpha_{1}$ and $\alpha_{2}$ are the SOC parameters in the spin-split bands along the $k_{x}$ direction, and $\beta$ is the SOC parameter in the spin-split bands along the $k_{y}$ direction [details derivation of the Hamiltonian $H_{Y}^{C_{s}}$, see the Appendix C]. 

The first and second terms of $H_{Y}^{C_{s}}$ in Eq. (\ref{4}) imposed the spin textures to exhibit the $S_{z}$ and $S_{y}$ spin components at $k_{x}\neq0$, while the $S_{x}$ spin component retains at $k_{x}=0$ due to the third term of the $H_{Y}^{C_{s}}$. As a result, the canted PST in the $y-z$ plane is expected at $k_{x}\neq0$, which matches well with the calculated spin textures obtained from the DFT calculation shown in Figs. 4(b) and 4(d). Moreover, the $H_{Y}^{C_{s}}$ leads to the energy dispersion, $E^{C_{s}}_{Y,\pm}(k)=E_{0}(k)\pm \sqrt{\alpha^{2} k_{x} + \beta^{2} k_{y}}$, where $\alpha=\sqrt{\alpha_{1}^{2}+\alpha_{2}^{2}}$ is the total SOC parameter defined in the spin-split bands along the $k_{x}$ direction. Our fitting calculation to the DFT band dispersion of the Ge$_{2}$SeTe monolayer around the $Y$ point at the VBM found that the calculated value of $\alpha$ parameter (3.10 eV\AA) is much larger than that of the $\beta$ parameter (0.008 eV\AA). The large different value between $\alpha$ and $\beta$ indicates that the band splitting is strongly anisotropic between $Y-M$ ($k_{x}$) and $\Gamma-Y$ ($k_{y}$) lines, which is consistent with the band dispersion shown in Fig. 3(b). In addition, the large value of $\alpha$ parameter implies that the large spin-splitting energy is achieved along $Y-M$ ($k_{x}$) line, which is also consistent with the spin-splitting energy obtained from the DFT calculations described in Fig. 3(d).   

We emphasized here that among the member of the 2D GeMC monolayers, the GeTe and Ge$_{2}$SeTe monolayers have the largest SOC parameter ($\alpha$). Importantly, compared to the other PST systems, the calculated value of $\alpha$ in both the monolayers is also the largest of all known PST materials so far [see Table 2]. Moreover, the emergence of the PST in these monolayers leads to the spatially periodic mode of the spin polarization, forming the persistent spin helix (PSH) mode with the wavelength of $l_{PSH}=\pi\hbar^{2}/(m_{x}^{*}\alpha)$ \cite{Bernevig}. We can estimate $l_{PSH}$ by considering $m_{x}^{*}$ and $\alpha$ parameters obtained from the bands along the $Y-M$ line, and find that the calculated value of  $l_{PSH}$ is about 6.53 nm and 8.52 nm for the GeTe and Ge$_{2}$SeTe monolayers, respectively. These values of $l_{PSH}$ are three-orders of magnitude smaller than that reported on semiconductor QW systems and comparable with that predicted on the 3D bulk PST system such as BiInO$_{3}$ (2.0 nm) \cite{Tao2018} and 2D PST systems including the doped SnSe monolayer (1.2 nm - 1.41 nm) \cite{Absor2019a} and the SnTe monolayers (1.82 nm - 8.8 nm) \cite{Absor2019b,Lee2020} [see Tabel 2]. Remarkably, the large SOC parameter ($\alpha$) and the small wavelength of the PSH mode ($l_{PSH}$) found in the present systems are important for miniaturization of spintronic devices operating at room temperatures.

Next, we discuss the tunability of the observed PST in the GeTe and Ge$_{2}$SeTe monolayers by introducing an in-plane strain. Here, we consider the strain created by changing the lattice parameters along $\vec{a}$ and $\vec{b}$ directions [see Fig. 1(a)]. We define degree of the strain as $\epsilon_{i=\vec{a},\vec{b}}=(a_{i}-a_{i_{0}})/a_{i_{0}}\times 100\%$, where $a_{i_{0}}$ and $a_{i}$ are the lattice parameters of equilibrium and strained structures, respectively, calculated along the selected $i=(\vec{a},\vec{b})$ direction. Although the electronic properties of the monolayers are sensitive to the strain, the PST preserves under large strain (up to $\pm 6$\%) due to the preserving of the crystal symmetry, thus we take it to show the relationship of the SOC parameter ($\alpha$) and the strain, which is plotted in Fig. 5(a). We find that stretching or compressing the monolayers significantly changes the magnitude of $\alpha$, i.e., the value of $\alpha$ sensitively increases (decreases) under the tensile (compressive) strains [Fig. 5(a)]. For an instant, in the case of the GeTe monolayer, when the tensile strain of $+6$\% is applied along the $\vec{a}$ ($\vec{b}$) direction, the value of $\alpha$ increases up to 6.65 eV\AA\ (5.75 eV\AA), which is much larger than the original value of 3.93 eV\AA. A similar trend also holds for the case of the Ge$_{2}$SeTe monolayer where the increasing value of $\alpha$ up to 4.57 eV\AA\ (4.02 eV\AA) is achieved under the tensile strain of $+6$\% along the $\vec{a}$ ($\vec{b}$) direction. Benefiting from the strong enhancement of $\alpha$ by the tensile strains, we should ensure that the wavelength $l_{PSH}$ of the PSH mode becomes significantly smaller than that of the original systems [see Fig. 5(b)], which is important for miniaturization spintronic devices.    

\begin{figure*}
	\centering	\includegraphics[width=1.0\linewidth]{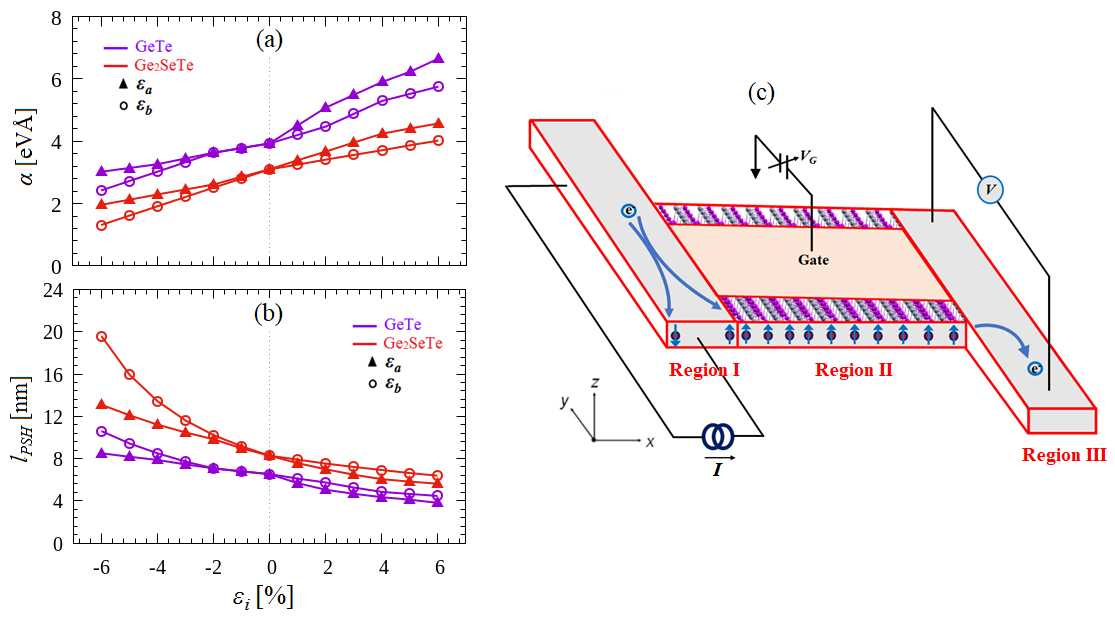}
	\caption{(a) The calculated SOC parameter, $\alpha$, and (b) the wavelength of the persistent spin helix (PSH) mode, $l_{PSH}$, as a function of strain $\epsilon_{i=\vec{a},\vec{b}}$ for the GeTe and Ge$_{2}$SeTe monolayers are shown. (c) Schematic view of the the spin Hall transistor device consisted of three regions: the left part (Region I) is the spin Hall injector region, the middle part (Region II) is the gate-controlled region, and the right part (Region III) is the inverse spin Hall detector region. }
	\label{fig:5}
\end{figure*}

Finally, based on the highly PST found in the present monolayers, we propose a spin Hall transistor (SHT) device as illustrated in Fig. 5(c). Motivated from the previous studies \cite{Choi2018, Jagoda}, we design the SHT device consisted of three parts, namely, region I, II, and III, representing an injection, gate-controlled, and detector regions, respectively. In region I, the pure spin currents polarized along the out-of-plane ($z$) orientation can move to the $x$-direction and are efficiently injected into the middle part of the device (region II) through the direct spin Hall effect (SHE). Subsequently, in region II in which the GeTe (or Ge$_{2}$SeTe) monolayer is take placed, the out-of-plane spin polarizations effectively induce the PSH mode in the crystal, which could be modulated by the out-of-plane electric field ($E_{z}$) driven by the gate electrode deposited on the top of the monolayers. Here, as a spin channel of the SHT, the minimum length of the spin channel in region II should be proportional to the wavelength of the PSH mode, $l_{PSH}$. For instant, a very short channel length of about 3.81 nm (5.62 nm) is achieved when the GeTe (Ge$_{2}$SeTe) monolayer with +6\% tensile strain along the $\vec{a}$ direction is used as a spin channel in the region II, which is much smaller than the channel length of the SHT reported on InAs QW system \cite{Choi2018}. Moreover, the presence of $E_{z}$ will determine the on/off logical functionality based on the preserving of the out-of-plane spin polarizations in the PSH mode. When $E_{z}=0$, the out-of-plane spin polarizations are robust to maintain the PSH mode, thus the spin polarizations are efficiently injected into region III without losing the spin information. By utilizing the inverse SHE effect in region III, the out-of-plane spin polarization is converted into the electric currents along the $y$ direction, which generates the Hall voltage. In contrast, the PSH mode is broken when $E_{z}\neq 0$ is applied [see Appendix D for the spin-resolved bands corresponding to their spin-splitting energy for the systems with an external out-of-plane electric field $E_{z}$], which results in the spin decoherence and significantly decreases the Hall voltage detected in the region III. 

We emphasized here that for the realization of the GeTe-based SHT device, the GeTe monolayer should be epitaxially grown on an insulator substrate. Consequently, the crystal symmetry of the GeTe monolayer will be further broken being similar to that of the Ge$_{2}$SeTe monolayer. In such a situation, the canted PST will maintain the SHT mechanism but with rather lower performance due to the admixtures of the in-plane spin polarization. This effect can be significantly reduced if the GeTe monolayer grown on the insulator substrate has a very small crystal mismatching effect and exhibits a highly ordered structure. Recently, several members of the 2D group IV monochalcogenide family such as SnTe and SnSe monolayers have been successfully synthesized on a properly graphitized SiC substrate \cite{KChang, KChang2020, KChang2019}. Considering the fact that the GeTe monolayer has the same symmetry as the SnTe and SnSe monolayers, the application of the graphitized SiC substrate for the GeTe-based SHT device is also highly plausible.

\section{Conclusion}

In summary, we have predicted the emergence of the PST in the 2D GeMC monolayer by performing first-principles DFT calculations combined with the symmetry analysis. We have studied the two stable formations of 2D GeMC monolayers, namely the pure Ge$X$ and Janus Ge$_{2}XY$ germanium mochalcogenides ($X, Y$: S, Se, and Te) monolayers and confirmed that the PST is observed around the VBM where the spin orientation is imposed by the lower point group symmetry of the crystal. In the case of the pure Ge$X$ monolayers, we have found that the PST is characterized by the fully out-of-plane spin polarization, which is protected by $C_{2v}$ point group symmetry, while the PST is canted in the $y-z$ plane for the case of the Janus Ge$_{2}XY$ monolayers due to the lowering symmetry into $C_s$ point group. More importantly, we have found that the PST sustains in the spin-split bands exhibiting large SOC parameters, which could be effectively controlled by applying the in-plane strains. The larger SOC parameter observed in the present systems leads to the fact that the smaller wavelength of the PSH mode is achieved, which is useful for miniaturization of the spin channel in the spin Hall transistor devices. 

Since the PST found in the present study is solely dictated by the $C_{2v}$ and $C_{s}$ point group symmetries of the crystal, we expect that the similar features are also shared by other materials having similar symmetry. Recently, there are a few other 2D materials that are predicted to maintain the similar symmetry of the crystals, including the 2D ferroelectric Ga$XY$ ($X$ = S, Se, Te; $Y$ = Cl, Br, I) family\cite{Absor2021} and 2D single-elemental multiferroic materials such as Te, and Bi \cite{Xiao, Pan}. Therefore, we have expected that our predictions will stimulate further theoretical and experimental studies in the exploration of the PST systems in the 2D-based materials, thus broadening the range of the 2D materials for future spintronic applications.

\appendix
\section{Symmetry analysis for the band degeneracy along the $\Gamma-Y$ line in the Ge$X$ monolayers}

In this appendix, we discuss the origin of the double degeneracy in the band structures of the Ge$X$ monolayers along the $\Gamma-Y$ line in the presence of SOC by considering the symmetry of the wave vector, $\vec{k}$. The wave vector $\vec{k}$ along the $\Gamma-Y$ line is invariant under $\bar{C}_{2y}$ screw rotation and $\bar{M}_{xy}$ glide mirror reflection. Both the symmetry operations hold the following algebra:
\begin{equation}
\label{1} 
\bar{M}_{xy} \bar{C}_{2y}=-e^{-ik_x}\bar{C}_{2y}\bar{M}_{xy}
\end{equation}
where the minus sign is originated from the fact that two spin rotation operators  $\sigma_{y}$ and $\sigma_{z}$ are anti-commutative, $\left\{\sigma_{y},\sigma_{z}\right\}=0$, so that both $\bar{C}_{2y}$ and $\bar{M}_{xy}$ operators are also anti-commutative, $\left\{\bar{C}_{2y},\bar{M}_{xy}\right\}=0$, on the $\Gamma-Y$ line.  As a result, for an eigenvector $\left|\psi_{m}\right\rangle$ of  $\bar{M}_{xy}$ operator with the eigenvalue of $m$, we obtain the following relation,
\begin{equation}
\label{2} 
\bar{M}_{xy}(\bar{C}_{2y}\left|\psi_{m}\right\rangle)=-m(\bar{C}_{2y}\left|\psi_m\right\rangle).
\end{equation}
The Eq. (\ref{2}) shows that $\left|\psi_{m}\right\rangle$ and $\bar{C}_{2y}\left|\psi_{m}\right\rangle)$ are different states with the same eigenvalue, thus ensures the double degeneracy along the $\Gamma-Y$ line.

\section{Spin textures at the VBM along the $\Gamma-Y$ line}

\begin{figure}
	\centering	\includegraphics[width=0.8\linewidth]{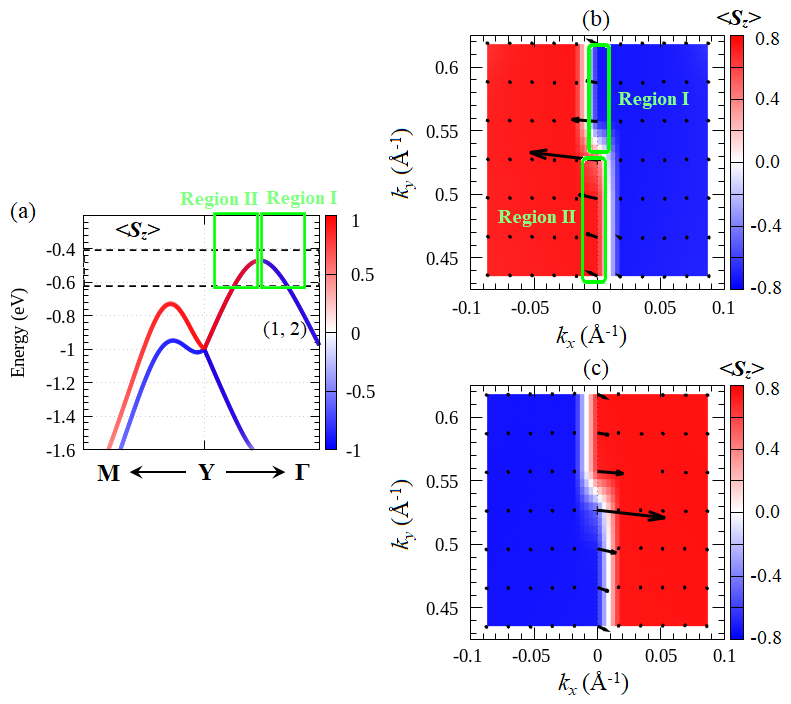}
	\caption{(a) $S_{z}$ spin projection of the bands of the GeTe monolayer at the VBM is shown, where the switching spin components along the $Y-\Gamma$ line are highlighted by the green line. The spin textures at the VBM at energy ranging from -0.6 eV to -0.4 eV at the VBM [as indicated by black-dashed lines in Fig. B1(a)] calculated for: (c) band (1) and (d) band (2). The arrows show the in-plane components of the spin, while the color indicates the out-of-plane component of the spins, $S_{z}$.}
	\label{fig:6}
\end{figure}

In this section, we present the calculated spin textures of the spin-split bands at the VBM to show the ensemble average values of the spin components along the $Y-\Gamma$ lines. Here, we calculate the spin textures at an energy ranging from -0.6 eV to -0.4 eV at the VBM [Fig. B1(a)]. The spin textures for the degenerate bands along the $Y-\Gamma$ lines are shown in Figs. B2(b) and B2(c) for the band (1) and (2), respectively. We find that the switching of the out-of-plane $S_{z}$ spin components are clearly observed along the $Y-\Gamma$ when crossing at $k_{y}$=0.52$ \AA^{-1}$. This switching is opposite between state (1) and (2) which has equal populations between $S_{z}$ and $-S_{z}$ spin components. Such switching of the spin polarization is consistent with the spin-resolved bands shown in Fig. 4(c). Moreover, we also identified the in-plane $S_{x}$ spin components around $k_{y}$=0.52$ \AA^{-1}$ as indicated by the black arrows in Figs. B1(b) and B1(c), which is also in agreement with the calculated $S_{x}$ spin projection of the bands presented in Fig. 4(c). 

\section{Derivation of $\vec{k}\cdot\vec{p}$ Hamiltonian}

To clarify the observed spin-splitting and the spin textures around the $Y$ point, we derive an effective $\vec{k}\cdot\vec{p}$ Hamiltonian, which can be deduced from the symmetry considerations. Here, we assume that only the linear term with respect to wave vector $\vec{k}$ contributing to the SOC Hamiltonian. We construct the Hamiltonian by identifying all symmetry-allowed term such that the following relation is obtained\cite{Winkler}:
\begin{equation}
\label{5} 
\hat{O}^{\dag}\hat{H}(k)\hat{O} =\hat{H}(k),	
\end{equation}								
where $\hat{O}$ denotes all symmetry operations belonging to the little group of the wave vector around the $Y$ point supplemented by time reversal symmetry.

\begin{table}[ht!]
\caption{Transformation rules for crystal momentum $k$ and spin operator $\sigma$ under considered point-group symmetry operations. Time-reversal symmetry defined as $i\sigma_{y} K$, where $K$ is complex conjugation and $\sigma$ denotes Pauli matrices, revers both the momentum and spin. The point-group operations are defined as $C_{2y}=i\sigma_{y}$, $M_{yz}=i\sigma_{x}$, and $M_{xy}=i\sigma_{z}$. } 
\begin{indented}
\item[]
\begin{tabular}{c c c  } 
\hline\hline 
Symmetry operations & ($k_{x}$, $k_{y}$)  & ($\sigma_{x}$, $\sigma_{y}$, $\sigma_{z}$) \\ 
\hline 
 $C_{2y}$ &  ($-k_{x}$, $k_{y}$)   &  ($-\sigma_{x}$, $\sigma_{y}$, $-\sigma_{z}$)  \\     
 $M_{yz}$&  ($-k_{x}$, $k_{y}$)   &  ($\sigma_{x}$, $-\sigma_{y}$, $-\sigma_{z}$)   \\		 
 $M_{xy}$ &  ($k_{x}$, $k_{y}$)   &  ($-\sigma_{x}$, $-\sigma_{y}$, $\sigma_{z}$)    \\         
\hline\hline 
\end{tabular}
\end{indented}
\label{table:Table 3} 
\end{table}

For the case of the GeTe monolayer, the little group of $Y$ $k$-point is $C_{2v}$, comprising two mirror symmetry operations, $M_{yz}$ and $M_{xy}$, and one two-fold rotation $C_{2y}$ around the $y$-axis. Taking into account the transformation role listed in Table B1, the symmetry-allowed linear spin-momentum coupling can be expressed as
\begin{equation}
\label{6} 
\hat{H}_{Y}^{C_{2v}}\left(k\right)=E_{0}(k)+\alpha k_{x}\sigma_{z}.							
\end{equation}

In contrast, both $M_{xy}$ and $C_{2y}$ symmetry operations are broken for the case of the Ge$_{2}$SeTe monolayer. Therefore, the crystal structure of the Ge$_{2}$SeTe monolayer belongs to the $C_{s}$ point group. Accordingly, the little point group at the $Y$ $k$-point also belongs to $C_{s}$ point group, which comprises only $M_{yz}$ mirror symmetry operation. The effective low-energy Hamiltonian around the $Y$ point in the Ge$_{2}$SeTe monolayer can again be deduced by considering only the $M_{yz}$ mirror symmetry operation. By using transformation role given in Table B1 for $M_{yz}$ operation, we obtain the following $\vec{k}\cdot\vec{p}$ Hamiltonian,
\begin{equation}
\label{7}
\hat{H}_{Y}^{C_{s}}\left(k\right)=E_0(k)+\left(\alpha_{1}\sigma_{z}+\alpha_{2}\sigma_{y}\right)k_{x}+\beta k_{y}\sigma_{x}.
\end{equation}

\section{Spin-resolved of the bands under an external out-of-plane electric field}

\begin{figure}
	\centering	\includegraphics[width=1.0\linewidth]{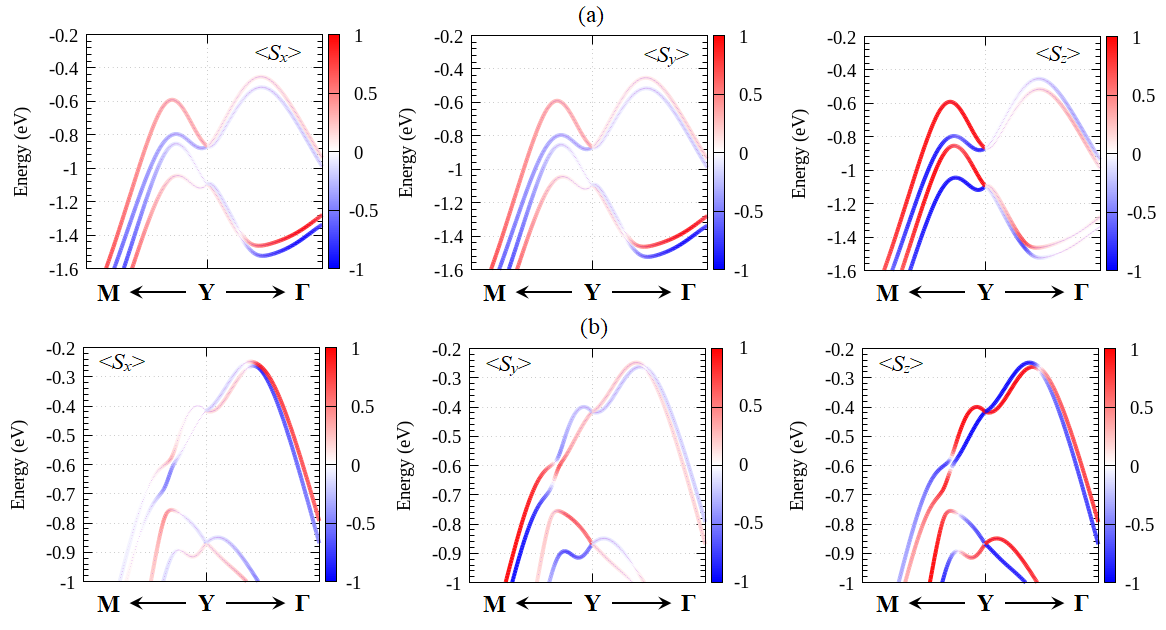}
	\caption{Spin-resolved projected to the bands at the VBM around the $Y$ point under an out-of-plane external electric field of 0.1 V/\AA\ for: (a) GeTe and (b) Ge$_{2}$SeTe monolayers}
	\label{fig:6}
\end{figure}

\begin{figure}
	\centering	\includegraphics[width=0.9\linewidth]{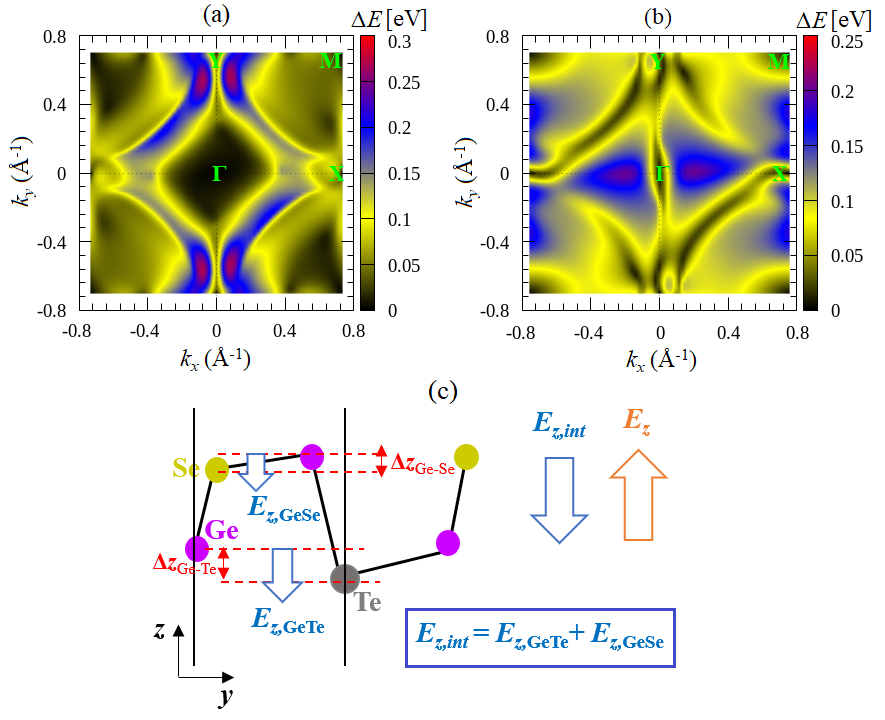}
	\caption{(a) and (b) Spin-splitting energy of the bands around the VBM mapped on the $k$-space in the first Brillouin zone for the GeTe and Ge$_{2}$SeTe monolayers, respectively, under an out-of-plane external electric field of 0.1 V/\AA. The color scales represent the magnitude of the spin-splitting energy similar to that shown in Figs. 3(a)-3(b). (c) Schematic view of the competition between the internal electric field $E_{z,int}$ and the applied electric field $E_{z}$ for the case of the Ge$_{2}$SeTe is shown. Here, the internal electric field $E_{z,int}$ is defined through the relation $E_{z,int}=E_{z,GeTe}+E_{z,GeSe}$, where $E_{z,GeTe}$ and $E_{z,GeSe}$ are the electric fields induced by the out-of-plane atomic distortion between the Ge and Te atoms ($\Delta z_{Ge-Te}$) and Ge and Se atoms ($\Delta z_{Ge-Se}$), respectively.}
	\label{fig:7}
\end{figure}

Figs. D1(a)-(b) show the calculated results of the spin-resolved projected to the bands at the VBM around the $Y$ point for the GeTe and Ge$_{2}$SeTe monolayers, respectively, under the influence of an external out-of-plane electric field ($E_{z}$) of  0.1 V/\AA. In the case of the GeTe monolayer, we find that the PST is broken by the electric field as indicated by the appearance of the in-plane $S_{x}$ and $S_{y}$ spin components in the spin-split bands [Fig. D1(a)]. Similarly, the canted PST in the case of the Ge$_{2}$SeTe monolayer is also broken by the $E_{z}$ since the in-plane $S_{x}$ and $S_{y}$ spin components appear in the spin-split bands [Fig. D1(b)].

The effect of the $E_{z}$ on the spin-splitting energy around the VBM is also shown in Figs. D2(a)-(b). In the case of the GeTe monolayer, it is clearly seen that application of the $E_{z}$ lifts the band degeneracy along the $\Gamma-Y$ line, where the spin-splitting energy up to 0.1 eV is achieved [Fig. D2(a)]. The emergence of the spin-splitting energy in the GeTe monolayer along the $\Gamma-Y$ line is due to the fact that both $M_{xy}$ and $C_{2y}$ symmetry operations are broken by the $E_{z}$. In contrast to the GeTe monolayer case, the spin-splitting bands of the Ge$_{2}$SeTe monolayer along the $\Gamma-Y$ line have already appeared without the $E_{z}$ [see Figs. 3(b) and 3(d)]. Introducing the $E_{z}$ in the Ge$_{2}$SeTe monolayer strongly decreases the spin-splitting energy along the $\Gamma-X$ and $Y-M$ lines as shown in Fig. D2(b). This is due to the fact that the competition between the internal electric field $E_{z,int}$ (induced by the out-of-plane atomic distortion) and the applied electric field $E_{z}$ decreases the net out-of-plane electric field as schematically shown in Fig. D2(c).

\ack{This work was supported by PDUPT (No.1684/UN1/DITLIT/DIT-LIT/PT/2021) and PD (No.2186/UN1/DITLIT/DIT-LIT/PT/2021) Research Grants funded by RISTEK-BRIN, Republic of Indonesia. This work was also partly supported by Grants-in-Aid for Scientific Research (Grant No. 16K04875) from JSPS and Grant-in-Aid for Scientific Research on Innovative Areas Discrete Geometric Analysis for Materials Design (Grant No. 18H04481) from MEXT Japan. The computation in this research was partly performed using the supercomputer facilities at RIIT, Kyushu University, Japan. Part of the computation in this research was performed using the computer facilities at Universitas Gadjah Mada, Republic of Indonesia.}

\section*{References}
\bibliography{Reference}

\end{document}